\documentclass[12pt,preprint]{aastex}

\slugcomment{Not to appear in Nonlearned J., 45.}
\shorttitle{$V$-band polarimetric separation of carbon stars}
\shortauthors{Goswami A.}
\begin{document}
\title{Evidence of $V$-band polarimetric separation of carbon stars  
at high Galactic latitude }
\author{Aruna Goswami\altaffilmark{ }, Sreeja S. Kartha\altaffilmark{ } }
\affil{Indian Institute of Astrophysics, Bangalore 560034, India\\
 e-mail: aruna@iiap.res.in
}

\and

\author{Asoke K. Sen\altaffilmark{ }}
\affil{Department of Physics, Assam University, Silchar, Assam, India\\}

\begin{abstract}
Polarization is an important indicator of stellar evolution, especially
for stars  evolving from red-giant stage to planetary nebulae.
However, not much is known about the  polarimetric properties of the
carbon-enhanced metal-poor (CEMP) stars, although they have been well 
studied in  terms of photometric as well as low- and high-resolution 
spectroscopy. We report here first-ever estimates of  $V$-band 
polarimetry of a group of CEMP stars.
  $V$-band polarimetry was planned as the $V$-band is known to show
maximum polarization among BVRI polarimetry for any scattering of 
light caused due to dust. Based on these estimates the program 
stars show a distinct classification into two: one with $p$\% $<$ 0.4 
and the other with $p$\% $>$ 1. Stars with circumstellar  material 
exhibit a certain amount of polarization that may be caused by
scattering of starlight due to circumstellar dust distribution into
non-spherically symmetric envelopes. The degree of polarization increases
with asymmetries present in the geometry of the circumstellar dust
distribution. Our results reflect upon these properties. While the 
sample size is relatively small, the polarimetric separation of the
two groups ($p$\% $<$ 0.4 and $p$\% $>$ 1) is very distinct; this finding,
therefore, opens up an avenue of exploration with regard to CEMP stars.
\end{abstract}


\section{Introduction}

A characteristic property of stars evolving from red giant stage  to
planetary nebulae (PNe) is polarization; observed polarization  reaches a
maximum  during  the proto-planetary stage and gradually decreases 
as the star evolves into a PN \citep{johnson91}. Polarimetric 
studies of post-asymptotic giant branch (AGB) stars by  \citet{trammell94} and 
\citet{parthasarathy93} have helped to identify 
aspherical morphology shells 
that originate during mass-loss phase  as common features of these objects.
Quasi-periodic variations in linear polarization are  known to be
observed in many late-type variables. However, not much work has been 
done on polarization of carbon stars; we report here  $V$-band 
polarimetric  measurements of a sample of carbon stars at high
Galactic latitude. Polarization of carbon stars are known to show
a flatter wavelength dependence than the oxygen-rich stars. 
Stars with circumstellar materials  exhibit  a certain
amount of linear polarization. The degree of polarization
increases with asymmetries present in the geometry of the shells.
By modeling the observed polarization,  it is possible to study
the geometry and structure of the  shells. Polarization models of 
circumstellar grain scattering  shows that  graphites do not
adequately explain the mean polarimetric behavior of carbon
stars \citep{raveendran91}. \citet{rowan83} have shown
that  graphites also do not adequately explain the IR spectra of 
circumstellar dust envelopes around carbon-rich objects. It was 
suggested  by \citet{czyzak82} that graphite formation in 
circumstellar envelopes is not a very likely process and its 
likely that carbon grains exist  most likely  as amorphous carbon.

While \citet{parthasarathy05} find no significant variation 
in polarization for HD~100746 at U, B, I bands, etc., \citet{raveendran91} 
finds mild  wavelength dependence of polarization in the case of a few 
carbon stars of Mira variables. In particular, the polarization was 
found to increase systematically toward  the red in RT~Pup,  
UU~Aur shows a dip in the polarization curve around the $V$ band, and  
X~Vel shows a larger polarization in the $R$ band than in the  $V$ band.
 Multiband 
polarimetric observations by \citet{serkowski75} showed that 
the interstellar polarization has a peak occurring at a median 
wavelength ${\lambda}$ = 0.545 ${\mu}$m. The maximum effect of 
interstellar polarization can be felt, therefore, in the yellow 
spectral region. It is  important therefore to undertake follow-up 
observations on several occasions to study the variations in 
polarization particularly for those that show significant 
polarization in the $V$ band. Our present sample includes
faint high latitude carbon stars  from \citet{christlieb01}. 
We have studied a large fraction of this sample based on
 low- as well as high-resolution spectroscopy (\citet{goswami05};
\citet{goswami06, goswami07, goswami10b}; \citet{goswami10a}). A primary 
characteristics 
of  these stars is that they provide evidence of nucleosynthesis of 
the  low- and intermediate-mass stars as most of them are polluted by
now-extinct AGB stars.  The basic parameters of the program stars 
are listed in Table 1.

\section { Observations and data reduction}
Observations were made using 2 m telescope of IUCAA, Pune using
the polarimeter (IMPOL) attached to it. The polarimeter available
at the cassegrain focus of 2 m IUCAA telescope has a rotating 
half-wave plate (HWP) and  Wollaston prism through which light passes
before forming a pair of images of an object on the CCD 
\citep{ramprakash98}.  The HWP can rotate in
several discrete steps, such that its fast axis makes angles
($\alpha$) with some reference direction (generally celestial
north-south). The light which is transmitted out  of the Wollaston
prism forms two images of any celestial source on the CCD, with
the ordinary and extraordinary set of rays. For each object
observations were taken at four different positions of the HWP:
0$^{0}$, 22$^{0}$.5, 45$^{0}$ and 67$^{0}$.5.

Among BVRI polarimetry, the $V$ band is known to show the maximum
polarization \citep{serkowski75} and is ideal for the detection of 
polarization caused due to scattering by dust. Moreover, because of 
a low photon flux in the $B$ band, the polarization measurements in the $B$ 
band often have large errors. For these reasons, we have  conducted
 only $V$-band polarimetry for the program stars.  In addition to the 
program stars  we have observed three polarization standard stars 
and one unpolarized standard star  for polarimetric calibration.

     Data reduction was carried out using various  tasks in IRAF. 
The task PHOT in APPHOT was used to measure the stellar flux. 
Zero polarization standard stars were observed to check for any 
possible instrumental error which proved to be smaller than  
0.1 \%. For stars which show low degree of polarization, measurement 
accuracy is an important issue. The measured errors are smaller 
than 0.1 \% for majority of the  observed stars.

\section{Results and Discussions}

 $V$-band  polarimetric measurements for the standard polarized
and unpolarized stars  are presented in Table 2. Estimates from 
literature  are also listed   for a comparison. A close agreement 
between the estimated and literature values  lends support to the 
reliability of our results. The estimated  error in polarization 
is less than or near about  0.1\% for all stars (Table 3) except 
for the star  HD~209621  for which the error is ${\pm}$ 0.23.
Since polarimetric error in our case is photon noise dominated,
 in the case of very bright stars this error  becomes close to zero. 
Our program stars include objects brighter as well as fainter  
than HD~209621. Thus, the error in HD~209621 can be considered 
to be a typical error value.

The polarization standard stars  HD~147084  and HD~160529,
both observed on the same night, show an offset of about 
${\sim}-$10$^{0}$;
the near equality of the offsets is as expected for consistency, and a
similar offset was also reported earlier  by \citet{sen00}.
The offset is essentially a result of constraints in instrumental
mounting, such as constraint in mounting the polarimeter with the
axis of  fixed HWP making as angle 0$^{0}$ with celestial north-south 
axis (this will give 0 offset).
We have applied an offset of -10$^{0}$ to the
estimated  position angles of all the stars observed on
the same night as that for HD~147084 and HD~160529.  The expected
final values of position angles ($\theta$$^{0}$), obtained  with
the necessary  offset corrections
are listed in  Table 3. 

 The linear polarization ($p$\%), with  the errors in the 
measurements of the program stars,  is presented  (Table 3) without 
any corrections applied to the observed values  for interstellar 
polarization. For stars with zero polarization (${\le}$0.1\%) the 
position angles are not defined, hence position angles are listed 
only for stars with non-zero polarization. The interstellar 
polarization is presumed to be negligible because of their high 
Galactic latitudes. The interstellar reddening, listed in Table 3, 
for the program stars is small.  For the program stars for which 
color excess have been known, it is instructive to compare the 
spatial distribution  of the color excess with polarization. In 
Figure 1 (upper panel), we  have plotted the $V$-band polarimetric 
estimates of the stars ($p$\%) against interstellar reddening $E(B-V)$
along with the mean interstellar polarization ${\sim}$ $3E(B-V)$; 
 \citep{parthasarathy05}.
Estimated  $V$-band polarization is found to be  larger than the 
mean interstellar contribution to the polarization in most of the 
cases indicating that they are intrinsically polarized.

Polarization does not seem to have any co-relation w.r.t. 
metallicity ([Fe/H]) of the stars (Figure 1, middle panel). The 
observed  $V$-band percentage polarizations are larger for high 
Galactic latitude objects (Figure 1,  lower panel);  however there 
are two outliers  (shown with triangles in Figure 1, lower panel) that are
high Galactic latitude objects but show weaker polarization. These two
objects  (HD~100764 and HE~1152-0355 with [Fe/H] ${\sim}$ $-$0.6 and $-$1.3,
respectively)  are not so metal-poor  compared to the other  stars 
that show larger polarization. It has been observed that the objects 
clearly fall  into two distinct groups; five objects showing
polarization below 0.4$\%$  and the other  five  showing  polarization
above 1.0$\%$. HD~206983 and HD~196944 in our sample do not
show significant  evidence of  linear polarization in the $V$ band.
 The properties of the individual stars are discussed  below.

\subsection {HD~206983, HD~196944} 
 HD~206983 is listed in the CH star catalog of 
\citet{bartkevicius96}. Recent high-resolution analysis of HD~206983 
by \citet{drake08} shows this object to be a Barium 
star with an  effective temperature of 4200 K  and a metallicity
of [Fe/H] = $-$1.43. Estimated $p$\% is 0.06$\pm$0.13. The $s$-process 
element-rich star HD~196944  is one of the first stars recorded to 
have shown very high  abundance of lead (Pb). This star, also  known  
as a lead star, is hotter (${\sim}$ 5200 K) and more metal-poor 
([Fe/H] = $-$2.25) than HD~206983 (\citet{aoki02}; \citet{van01}). 
Both  C and N abundances are enhanced 
 ([C/Fe] = 1.2, [N/Fe] = 1.3). This is a high-velocity object with 
estimated radial velocity $-$174.76 $\pm$ 0.36 km s$^{-1}$,
 the abundance distribution observed in this star is interpreted  
as resulting from now-extinct companion AGB star.
 HD~196944 is mentioned in SIMBAD as a star with envelop of CH;
the envelop might have resulted from as the companion star evolved 
through AGB and PN phases suffering mass loss.
 The estimated  polarization ($p$\% = 0.09$\pm$0.02) is not significant;
it is also likely that the  non-detection of polarization would perhaps 
mean that the CH envelop is  symmetrically structured. No previous 
polarization estimates are available for these two objects.\\

\subsection {HD~100764, HD~168986, HD~198269, HD~209621, HE~1152$-$0355}
  Estimated $p$\% for these objects are, respectively, 0.330$\pm$0.06, 
0.358$\pm$0.03, 0.269$\pm$0.02, 0.114$\pm$0.23, and 0.245$\pm$0.09.

Among  these objects only  HD~100764, a non-pulsating R-type 
carbon star, has  previous polarimetric measurements  in literature. 
Percentage  polarization estimates in the  $V$ band on  two occasions by 
\citet{parthasarathy05} are, respectively,  0.20$\pm$0.05 (JD 2,447,971) 
and 0.42$\pm$0.14, (JD 2,448,398). Multiband photo-polarimetry  of 
HD~100764  shows no  significant variations in $p$\% in $B$, $V$, $U$ and $I$ 
bands. No information is available on  time variability of 
polarization for this star. Our estimated $p$\% = 0.33$\pm$0.06  
is not too different from the previous estimates.

The reddening in the direction of HD~100764 ($E(B-V)$ = 0.02; Eggen 1972)
 is consistent with the high Galactic latitude of the star.
The observed $V$-band polarization is greater than the mean 
interstellar contribution to the  polarization, indicating that the 
star is intrinsically polarized.  \citet{richer75} derived the  absolute
visual magnitude  of  $M_{v}$ = $+0.3$ for this object  that corresponds 
to a distance of 500 pc with  $E(B-V)$ = $0.02$. From a detailed 
spectroscopic analysis  \citet{dominy84} found $T_{eff}$ = 4850 K, 
log\,$g$ = 2.2 and a  metallicity [Fe/H] = $-$0.6 for this object.

HD~100764 shows significant far-infrared excess fluxes at 12${\mu}$m, 
25${\mu}$m, 60${\mu}$m and 100${\mu}$m with uncertainties of 4\%, 
8\%, 12\%, and 12\%, respectively (\citet{beichmann85}).
The silicate dust shell around HD~100764 is believed to absorb the
radiation from the star and re-emit the radiation in the infrared.
Modeling of its circumstellar dust by \citet{skinner94} revealed  a
massive dusty disk. 
 \citet{parthasarathy91} suggested    that HD~100764 is a
star  with a detached cold dust shell and that the optical region
fluxes of HD~100764 do not show evidence for significant
reddening, and argued  that the dust shells are  optically
thin with relatively large  dust grain size  and dust is
confined in the form of thin disc around the star and hence shows
less reddening. The silicate dust shells around  this star may be
the result of mass loss experienced by the  star during core
helium flash, which has taken place recently. The presence of dust
shell around  HD~100764 also  suggests that it has experienced
significant mass loss in the recent past.

 The star HD~168986  is listed in the Barium star catelog of 
\citet{lu91}.  Barium stars are however not known to have dust 
shells or discs around them \citep{dominy86}. This star 
mentioned as a peculiar star in SIMBAD shows an estimated
$p$\% =  0.36$\pm$0.03, that seems to represent the  intrinsic
polarization property of the star. 
No previous polarization estimates are available for this object.

HD~198269 is listed in the CH star catalog of \citet{bartkevicius96}.
In SIMBAD this star is mentioned as a star with envelop of CH.
\citet{lee74} gave an upper limit to its distance estimate as 
 ${\sim}$  750 pc that corresponds to  $M_{v}$ = $-$1.66, an upper 
limit to the absolute magnitude. This is a non-variable CH star
of type R with a mass  of about 0.8 $M_{\odot}$ \citep{wallerstein73}
and  is similar in chemical compositions in  comparison to  other CH stars.
With $E(B-V)$ ${\le}$ 0.07 mag for this star, the estimated 
$p$\% = 0.27$\pm$0.29 is marginally higher than the mean 
interstellar contribution to the polarization.
No previous polarization estimates are available for this object.

 HD~209621 and HE~1152-0355, both are   confirmed CH stars.  Surface 
composition of these two objects show abundance patterns that are 
consistent with the abundances generally noticed in CH stars, 
essentially arising from $s$-processing (\citet{goswami10a}; 
\citet{goswami06}).  HD~209621 shows  high abundances of $r$-process 
element  Eu as well as of the third-peak $s$-process element Pb. 
Eu and Pb are, however,  not detected in the spectrum of
HE~1152-0355. Both these objects show larger enhancement of the 
second-peak $s$-process elements  as compared to those of the first-peak 
$s$-process elements. Both are high-velocity objects. HD~209621 is  
a radial velocity variable with a period of 407.4 days 
\citep{mcclure90}; however, radial velocity variability 
is not yet confirmed for HE~1152-0355. Our $V$-band polarimetric 
estimates  indicate no significant polarization from HD~209621, and 
HE~1152-0355 shows polarization that is marginally  above the mean
interstellar contribution to the polarization.\\

\subsection {HE~1027$-$2501, HE~1305$+$0007, HE~1429$-$0551, LP~625$-$44,
HE~1523$-$1155}
These objects show significant polarization with $p$\% ${\ge}$ 1,
except for HE~1523-1155 for which we have obtained $p\%$ = 0.849
$\pm$ 0.08. Reddening estimates $E(B-V)$ range from 0.03 to 0.08 for these stars;
the values of $E(B-V)$ for the HE stars are taken from \citet{beers07}.

A low carbon isotopic ratio ($^{12}$C/$^{13}$C $<$ 10) for 
HE~1027-2501  indicates  that the star is on the first ascent of 
the giant branch wherein  the material transferred from the now 
unseen companion has been mixed into the CN-burning region of the 
CH star or constitute a minor fraction of the envelop mass of 
the CH star. Such low values are  believed to be due to convection 
which dredges up the products of internal CNO cycle to the stellar 
atmosphere in the ascending red giant branch (RGB). When the star 
reaches the AGB stage,  fresh $^{12}$C may
 be supplied from the internal He-burning layer to the stellar surface,
 leading to an increase of $^{12}$C/$^{13}$C ratio.

HE~1305$+$0007 shows enhancement of both $r$- and $s$-process elements 
including lead. The second-peak $s$-process elements are  more 
enhanced than the first-peak $s$-process elements. This is a 
low-metallicity and  high-velocity object 
($V_{r}$ = $+$217.8 $\pm$ 1.5 km s$^{-1}$); its atmospheric 
parameters are consistent with a present location on the 
RGB (\citet{goswami06}).

The star LP~625$-$44  is a carbon- and $s$-process-element-rich  
very metal-poor subgiant.  Abundance estimate derived  using  
the O $I$ triplet around  7770 \AA\, shows excess of oxygen by
a factor of 10 (Aoki et al. 2002); and  Na enhancement by about 
a factor of 50 in comparison to HD~140283 (a metal-poor subgiant 
with normal abundance ratio)  and a high Mg abundance 
 ([Mg/Fe] = 1.12 $\pm$ 0.24).  High abundance of Na suggests,  that
hydrogen burning in the $^{22}$Ne-rich layer in an AGB
 star must have   produced the abundance pattern of this object.
The Pb enhancement  shown by LP~625$-$44  is not high enough to be
placed in  the group of  lead stars. The abundance ratio of $s$-process 
elements at the second peak (La, Ce, and Nd) to that at the third 
peak (Pb) in LP 625-44 is significantly higher (by a factor of 5) than 
that in the  $s$-process element-rich lead star  HD~196944
 (\citet{van01}).
Unlike CH stars, the radial velocity of this object is low 
(${\sim}$ 30 km s$^{-1}$) and the variation of the radial velocity  is 
expected to be  for about 200 days \citep{aoki00}.

Like LP~625-44, HE~1429-0551, and HE~1523-1155 are also low-velocity 
objects
(${\sim}$ $-$44.9 and ${\sim}$ $-$46.03 km s$^{-1}$, respectively).
 Both   show high abundances of C, N, and Mg relative to Fe. Estimated  
carbon isotopic ratio of $^{12}$C/$^{13}$C for HE~1429-0551 is high
 $\sim ~30^{+20}_{-10}$ \citep{aoki07}. Such high ratios are 
generally noticed in C-N stars. With a marginal difference in the
 molecular band depths, the spectra of the star HE~1523-1155 
closely resemble the spectrum of HD~5223, a  well-known CH giant
 \citep{goswami05}. Both the HE stars  show high  barium abundance 
with respect to Fe ([Ba/Fe] = 1.57 and 1.72, respectively), and 
log($L/L_{\odot}$) are, respectively,  2.56 and 2.50 \citep{aoki07}.
The observed $V$-band polarization estimates are markedly higher 
than the mean  interstellar contribution to the polarization and 
seem to represent the stars intrinsic polarization properties.

\section{Conclusions}
Our  $V$-band polarimetric observations presented for a sample 
of 10 carbon-enhanced metal-poor stars show  five objects with 
polarization $p$\% $>$ 1 and the other  five  with $p$\% $<$ 0.4. 
While the size of the sample is relatively small, it should be 
noted that these stars belong to similar group in terms of carbon 
enhancement and metal deficiency. Thus, a clear separation of these
 objects into well-separated $V$-band polarimetric groups is significant.
Intrinsic polarization of late-type stars are  usually attributed
either to the Rayleigh scattering of  the light emerging from the 
limb of a star  in which case 
the sources
of photometric  asymmetry  necessary for a net observable
polarization are suggested to be non-radial pulsation of the star,
variation of temperature over the surface and the presence of
giant convection cells (\citet{harrington69}; \citet{schwarzschild75})
 or due to   scattering by molecules and dust grains in an 
extended asymmetric circumstellar envelope (\citet{kruszewski68}; 
\citet{shawl75}; \citet{daniel78}). It seems  circumstellar grain scattering 
is the main mechanism responsible for the observed polarization in
the sample of program  stars.  It would be worthwhile to
conduct polarimetric observations for the entire sample of carbon 
stars from Hamburg ESO survey \citep{christlieb01} for a 
statistical  evaluation of their polarization properties.

Except for HD~100764, our results are  the first   estimates of
polarization for the program stars.  While the separation of the 
stars into two distinct polarization groups is a significant 
feature, it is  not clear at the moment if this is a time-dependent 
property.  While a few of the CEMP stars are confirmed as 
radial velocity variables and binaries, binarity of others are 
not yet established. Two important means for establishing 
variability and binarity of stars are long-term photometric 
observations as well as long-term spectroscopic radial
velocity measurements. Time variation of polarization measurements
for  these objects would also  be an indicator of their  
variability caused by the morphology of the circumstellar dust 
clouds that might have originated during AGB  mass loss
of the companion stars. Thus, the detection of polarization from 
these objects would lend  support to those  physical scenarios 
that consider binarity  of the objects  to explain their
formation mechanism and peculiar abundances of heavy elements. \\

{\it Acknowledgement}\\
 We thank  the staff at IUCCA observatory  for time allocation and 
assistance during the observations.  This work made use of the SIMBAD 
astronomical database, operated at CDS, Strasbourg, France, and the NASA 
ADS, USA.  We thank  Prof. A.  V. Raveendran for useful discussions.
Sreeja S. Kartha is a JRF in the  DST project SR/S2/HEP-09/2007;
funding from this project is greatfully acknowledged.
\\

\begin{figure*}
\includegraphics[angle=0,height=15cm,width=15cm]{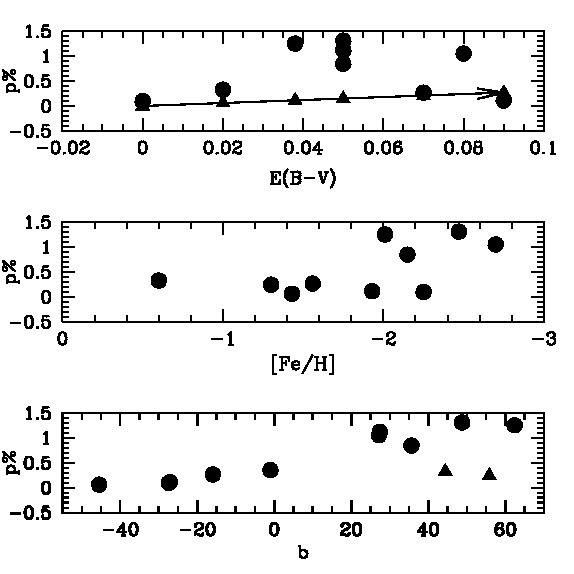}
\caption{ Upper panel: observed $V$-band percentage polarization for the 
program stars
vs. interstellar $E(B-V)$. The solid line represents the mean interstellar
polarization (${\sim}$ $3E(B-V)$).  
 Middle panel: observed $V$-band percentage polarization of the program stars
w.r.t. their metallicities [Fe/H]. Polarization does not
seem to have any co-relation w.r.t. metallicities  of the stars.
Bottom panel:  the observed $V$-band percentage  polarizations of the program
stars are plotted w.r.t. their Galactic latitudes.
Estimated  polarizations are larger for  objects with positive Galactic 
latitudes; however, there are two outliers  (shown with triangles) that  show 
weaker polarization. These two
objects  (HD~100764 and HE~1152-0355 with [Fe/H] ${\sim}$ $-$0.6 and $-$1.3, 
respectively)  are not so metal-poor  compared
to the other  stars that show larger polarization.}
\label{Figure 1}
\end{figure*}

{\footnotesize
\begin{table*}
{\bf Table 1: Basic  Parameters of the Program Stars}\\
\tiny
\begin{tabular}{ l c c c c c c c c c c c  }
              &            &           &        &       &    &       &       &      &     &  \\
\hline
Stars         &R.A.(2000)    & Decl.(2000) & $+l$      & $b$     & $V$   &  $J$  & $ H$ & $K_{s}$&  Date of Obs.  &\\
              &            &           &        &       &     &       &      &        &   & \\
\hline
              &            &           &        &       &     &       &      &        &   & \\
HE~1027$-$2501&10~29~29.5&$-$25~17~16.2&266.6829& +27.4161 &12.70&10.627& 9.896 & 9.722 & 2009 Apr 30 &  \\
HD~94851$^*$      &10~56~44.2&$-$20~39~52.6&269.8196& +34.7281 & 9.27 & 8.869 & 8.816 & 8.771 & 2009 Apr 30 &     \\
HD100764      &11~35~42.7&$-$14~35~36.6&276.8586& +44.4107  &8.73 &7.048 & 6.60 & 6.513  & 2009 Apr 30 &  \\
HE~1152$-$0355&11~55~06.1&$-$04~12~24.0&277.3246& +55.8385  &11.43&9.339 & 8.665 & 8.429 & 2009  Apr 30&  \\
HE~1305$+$0007&13~08~03.8&$-$00~08~47.4&311.9430&  +62.4340 &12.22&10.247& 9.753 & 9.600 & 2009  Apr 30&  \\
HE~1429$-$0551&14~32~31.3&$-$06~05~00.2&343.0186& +48.7605   &12.60&10.734& 10.272&10.066 & 2009  Apr 30&  \\
HE~1523$-$1155&15~26~41.0&$-$12~05~42.6&351.8699&  +35.6301 &13.22&11.372& 10.846& 10.748& 2009  Apr 30&  \\
HD~147084$^*$     &16~20~38.1&$-$24~10~09.5&352.3279& +18.0503& 4.55 & 2.222 & 1.899 & 1.681 & 2009 Apr 30&   \\
LP~625$-$44      &16~43~14.0&$-$01~55~30.2&015.1463& +27.1749  &11.85&10.432&10.058 & 10.825& 2009 Apr 30 &   \\
HD~160529$^*$     &17~41~59.0&$-$33~30~13.7&355.7021&  -01.7322&6.77 &3.547 & 3.056 & 2.790 &  2009  Apr 30&   \\
HD~168986     &18~23~10.1&$-$15~36~36.5&015.8506& -00.9879&9.21 &6.901 & 6.379 & 6.184 & 2009 Apr 30 &  \\
HD~196944     &20~40~46.0&$-$06~47~50.6&039.5635& -27.3703  &8.41 &7.017 & 6.626 & 7.504 & 2009 Apr 30 &  \\
HD~198269      &20~48~36.7&$+$17~50~23.7&063.2273& -15.9516  &8.12 &6.078 & 5.505 & 5.385 & 2009  Apr 30&  \\
HD~206983     &21~46~09.3&$-$15~14~39.7&038.6462& -45.4725&9.45 &6.934 & 6.247 & 6.035 & 2009  Apr 28&   \\
HD209621      &22~04~25.1&$+$21~03~08.9&078.6453& -27.1175    &8.86 &6.661 & 6.045 & 5.913 & 2009 Apr 28 &  \\
              &          &             &        &       &      &      &       &       &          \\
\hline
\end{tabular}

{\bf Note.} The stars marked with $*$ are polarization  standard stars.
\end{table*}
}

{\footnotesize
\begin{table*}
{\bf Table 2: $V$-band Polarimetry of Polarization Standard Stars}\\
\tiny
\begin{tabular}{ | c | c | c | c | c | c | c | }
\hline
Star Names     &  HJD               & $p$\%           & $\theta$$^{0}$ & $p$\%             & $\theta$$^{0}$  & Remarks \\
(1)            &  (2)               & (3)$^*$         &  (4)$^*$       &   (5)       &   (6)        & (7)     \\
\hline
               &                    &                 &             &   &   & \\
HD~94851       & 2454952.13280      & 0.11 $\pm$0.12  &             & 0.057 $\pm$ 0.018  &   &  \citet{turnshek90} \\
               &                    &                 &             & 0.134      &   &  \citet{rautela04}\\
               &                    &                 &             & $\le$ 0.1  &  &   \citet{leonard05}\\
               &                    &                 &             &   &   & \\
               &                    &                 &             &   &   & \\
HD~147084      & 2454952.25726     & 4.12 $\pm$0.02  &  32.2        & 4.16 $\pm$ 0.01  & 32.1 $\pm$1.9 & \citet{mcdavid99} \\
               &                    &                 &             & 4.17 $\pm$ 0.008 & 32.9   &  \citet{turnshek90}\\
               &                    &                 &             & 4.19 $\pm$ 0.03  &  32.4 $\pm$ 0.9 & \citet{chavero06} \\
               &                    &                 &             & 4.23           &  33.5 &   \citet{clarke98} \\
               &                    &                 &             & 4.21  &  31.3 &  \citet{weitenbeck04} \\
               &                    &                 &             &   &   & \\
HD~160529      & 2454952.46513     & 7.45 $\pm$0.02   & 20.9      &7.35 $\pm$ 0.55 &   &  \citet{reiz98} \\
               &                    &                 &             & 7.3 $\pm$ 0.54 & 20  &  \citet{serkowski74}\\
               &                    &                 &             & 7.52           & 20.1  &   \citet{clarke98} \\
               &                    &                 &             &   &   & \\
\hline
\end{tabular}

{\bf Notes.} Estimates  listed  in Columns 5 and 6 are  from references 
listed in Column 7.\\
$^*$Our estimates of $p$\% and offset corrected position angles (magnitudes)\\
\end{table*}
}

{\footnotesize
\begin{table*}
{\bf Table 3: $V$-band  Polarimetry of Carbon Stars }\\
\tiny
\begin{tabular}{ l c c c c c c  c  c  }
\hline
               &       &           &        &        &         &                &               &       \\
Star Names     & $E(B-V)$& $T_{eff}$ & log$g$ & [Fe/H] & Reference    &   HJD     & p$\%$         & $\theta$$^{0}$  \\
\hline
               &       &           &        &        &         &                &               &       \\
HD~209621      & 0.09 & 4500 & 2.0 & -1.93  & 1   & 2454950.47784     & 0.11 $\pm$ 0.23  &             \\
HD~206983      & ---  & 4200  & 1.4 & -1.43   & 2   & 2454950.48532      & 0.06 $\pm$ 0.13  &             \\
HD~100764      & 0.02 &4850  & 2.2 & -0.6  & 3   & 2454952.14664      & 0.33 $\pm$ 0.06  & 4.0  \\
HE~1152$-$0355 & 0.026& 4000 & 1.0 & -1.3  & 4   &  2454952.15514     & 0.24 $\pm$ 0.09  & 30.5   \\
HE~1027$-$2501 & 0.05 & ---  & --- & ---   & -   & 2454952.18021     & 1.13 $\pm$ 0.11  &  12.4  \\
HE~1305$+$0007 & 0.038& 4750 & 2.0 & -2.01  & 4    &  2454952.22191     & 1.26 $\pm$ 0.04  & 41.9  \\
HE~1429$-$0551 & 0.05 & 4700 & 1.5 & -2.47  & 5    & 2454952.27449     & 1.31 $\pm$ 0.06  & 28.9  \\
LP~625$-$44    & 0.08 & 5500 & 2.5 & -2.70  & 6    & 2454952.31728     & 1.06 $\pm$ 0.03  & 8.0  \\
HE~1523$-$1155 & 0.05 & 4800 & 1.6 & -2.15  & 5    & 2454952.35635      & 0.85 $\pm$ 0.08  & 30.3    \\
HD~168986      & ---  & ---  & --- & ---   & -    & 2454952.39775      &0.36 $\pm$ 0.03   & 32.5  \\
HD~196944      & 0.0  & 5250 & 1.8 & -2.25  & 7  & 2454952.47013     & 0.09 $\pm$ 0.02   &           \\
HD~198269      & $ <$ 0.07& 4460 & --- & -1.56  & 8 & 2454952.47835     & 0.27 $\pm$ 0.02   & 39.5 \\
               &       &             &                 &          \\
\hline
\end{tabular}

References. (1) \citet{goswami10a}; (2) \citet{drake08}; 
(3) \citet{dominy84}; (4) \citet{goswami06}; \\
(5) \citet{aoki07}; (6) \citet{aoki02}; (7) \citet{van01};
  (8) \citet{lee74}.\\
\end{table*}
}
\end{document}